\documentclass[ALICE,manyauthors]{cernphprep}

\usepackage[comma,square,numbers,sort&compress]{natbib}
\usepackage{hyperref}
\usepackage{lineno}

\def\pt{\mbox{$p_{\rm T}$}}   
\def\snntwo{\mbox{$\sqrt{s_{\rm NN}} = 2.76$}} 
\def\snnfive{\mbox{$\sqrt{s_{\rm NN}} = 5.02$}}   
\def\pb {Pb--Pb }
\def\au {Au--Au }
\def\pp {$pp$}
\def\j {\mbox{J/$\psi$ }}
\def\vv {$v_2$ }
\def\vpt{\mbox{$v_2(p_{\rm T})$}}
\def\raa {\mbox{$R_{\rm{AA}}$}}

\newcommand{\snn}{\ensuremath{\sqrt{s_{\rm NN}}}}
\newcommand{ \be }{\begin{eqnarray}}
\newcommand{ \ee }{\end{eqnarray}}

\newcommand{\dEdx}{\ensuremath{\text{d}E/\text{d}x}\xspace}

\usepackage[sort&compress,numbers]{natbib}
\usepackage{doi}
\usepackage{hyperref}
\usepackage{lineno}
\usepackage{multirow}
\usepackage{bm}

\usepackage{amssymb}
\usepackage{amsmath}

\begin{document}%
%

%
\begin{titlepage}
\PHyear{2017}
\PHnumber{237}                 
\PHdate{11 September}              
%
%
\title{\j elliptic flow in \pb collisions at $\mathbf{\sqrt{s_{\rm NN}} = 5.02}$ TeV}
\ShortTitle{\j elliptic flow in \pb collisions at $\sqrt{s_{\rm NN}} = 5.02$ TeV}   
%
\Collaboration{ALICE Collaboration\thanks{See Appendix~\ref{app:collab} for the list of collaboration members}}
\ShortAuthor{ALICE Collaboration} 
\begin{abstract}

We report a precise measurement of the \j elliptic flow in \pb collisions at \mbox{$\sqrt{s_{\rm NN}}~=~5.02 $~TeV} with the ALICE detector at the LHC. 
The J/$\psi$ mesons are reconstructed at mid-rapidity ($|y| < 0.9$) in the dielectron decay channel and at forward rapidity ($2.5<y<4.0$) in the dimuon channel, both down to zero transverse momentum.
At forward rapidity, the elliptic flow \vv of the \j is studied as a function of transverse momentum and centrality. 
A positive \vv is observed in the transverse momentum range $2 < p_{\rm T} < 8$ GeV/$c$ in the three centrality classes studied and confirms with higher statistics our earlier results at $\sqrt{s_{\rm NN}} = 2.76$ TeV in semi-central collisions.
At mid-rapidity, the \j \vv is investigated as a function of transverse momentum in semi-central collisions and found to be in agreement with the measurements at forward rapidity. These results are compared to transport model calculations. 
The comparison supports the idea that at low \pt\  the elliptic flow of the \j originates from the thermalization of charm quarks in the deconfined medium, but suggests that additional mechanisms might be missing in the models.
\end{abstract}
\end{titlepage}
\setcounter{page}{2}
%

Extreme conditions of temperature and pressure created in ultra-relativistic heavy-ion collisions enable exploration of the phase diagram region where Quantum Chromodynamics (QCD) predicts the existence of a deconfined state, the Quark-Gluon Plasma (QGP)~\cite{Bjorken:1982qr,Aoki:2006we}. 
Heavy quarks are produced through hard-scattering processes prior to the formation of the QGP and experience the evolution through interactions in the medium. 
Therefore, the measurement of bound states of heavy quarks, such as the J/$\psi$, is expected to provide sensitive probes of the strongly-interacting medium~\cite{Andronic:2015wma}.
Theoretical calculations based on lattice QCD predict a \j suppression to be induced by the screening of the color force in a deconfined medium which becomes stronger as the temperature increases~\cite{Matsui:1986dk,Digal:2001ue}. 
In a complementary way to this static approach, \j suppression can be also interpreted as the result of dynamical interactions with the surrounding partons~\cite{Ferreiro:2012rq,Zhao:2011cv,Zhou:2014kka}.
Within these scenarios, the \j suppression, experimentally quantified via the nuclear modification factor, \raa\ (the ratio between the yields in \pb to \pp\ collisions normalised by the number of nucleon–-nucleon collisions), is expected to become stronger (smaller \raa) with higher initial temperatures of the QGP, hence with higher collision energies. 
However, the \raa\ of inclusive~\footnote{Inclusive \j include prompt \j (direct and decays from higher mass charmonium states) and non-prompt \j (feed down from b-hadron decays). In this Letter, all \j measurements refer to inclusive \j production unless otherwise stated.} \j with transverse momentum $\pt < 8$ GeV/\textit{c} observed by the ALICE Collaboration in \pb collisions at \snntwo~ TeV~\cite{Abelev:2012rv} and \snnfive~TeV~\cite{Adam:2016rdg} is larger than what has been measured at lower energies at the Relativistic Heavy Ion Collider (RHIC)~\cite{Adare:2006ns,Adare:2011yf,Adamczyk:2013tvk,Adamczyk:2016srz} and exhibits almost no centrality dependence. 
Furthermore, in central collisions the measured \raa\ values decrease from low to high \pt\ \cite{Adam:2015isa,Abelev:2013ila}.
The \j \raa\ enhancement from RHIC to LHC energies can be explained by theoretical models~\cite{BraunMunzinger:2000px, Andronic:2011yq, Zhao:2011cv, Liu:2009nb, Zhou:2014kka, Ferreiro:2012rq} which include a dominant contribution from \j (re)generation through (re)combination of thermalized charm quarks in the medium, during or at the phase boundary of the deconfined phase~\footnote{The terms (re)generation and (re)combination denote the two possible mechanisms of \j generation by combination of charm quarks at the QGP phase boundary and the continuous dissociation and recombination of charm quarks during the QGP evolution.}.

Additional observables are required to better constrain theoretical models and study the interplay between suppression and regeneration mechanisms~\cite{Liu:2009gx}.
The azimuthal anisotropy of the final-state particle momentum distribution is sensitive to the geometry and the dynamics of the early stages of the collisions.
The spatial anisotropy in the initial matter distribution due to the nuclear overlap region in non-central collisions is transferred to the final momentum distribution via multiple collisions in a strongly coupled system~\cite{Ollitrault:1992bk}. 
The beam axis and the impact parameter vector of the colliding nuclei define the reaction plane. 
The second coefficient ($v_2$) of the Fourier expansion of the final state particle azimuthal distribution with respect to the reaction plane is called elliptic flow.

Within the transport model scenario~\cite{Zhao:2011cv, Liu:2009nb}, (re)generated \j inherit the flow of the (re)combined charm quarks. 
If charm quarks do thermalize in the QGP, then (re)generated \j can exhibit a large elliptic flow.
In contrast, only a small azimuthal anisotropy, due to the shorter in-plane versus out-of-plane pathlength, is predicted for the surviving primordial J/$\psi$.
The ALICE and CMS collaboration have measured a positive elliptic flow of D mesons in Pb-Pb collisions at \snnfive~TeV \cite{Acharya:2017qps,Sirunyan:2017plt}. The comparison of \j and D-meson \vv could help to constrain the dynamics of charm quarks in the medium and the theoretical model calculations ~\cite{He:2014cla,Du:2015wha,Lin:2003jy}.

At RHIC, the STAR Collaboration measured, in \au collisions at $\snn = 200$ GeV, a \j \vv consistent with zero, albeit with large uncertainties~\cite{Adamczyk:2012pw}.
At the LHC a first indication of positive \j \vv was observed by the ALICE Collaboration in semi-central \pb collisions at \snntwo~ TeV with a $2.7\sigma$ significance for inclusive \j with $2 < p_{\rm T} < 6$~GeV/$c$ at forward rapidity~\cite{ALICE:2013xna}. 
The CMS Collaboration also reported a positive \vv for prompt \j at high \pt\ and mid-rapidity~\cite{Khachatryan:2016ypw}.
A precision measurement of the \j \vv in \pb collisions at the highest LHC energy will provide valuable insights on the \j production mechanisms and on the thermalization of charm quarks. 
Indeed, the higher energy density of the medium should favor charm quark thermalization, and thus increase its flow. 
In addition, the larger number of produced $\rm{c\bar{c}}$ pairs should increase the fraction of \j formed by regeneration mechanisms, both leading to an increase of the observed \mbox{J/$\psi$} $v_2$.

In this Letter, we report ALICE results on inclusive \j elliptic flow in \pb collisions at \snnfive~ TeV for two rapidity ranges. At forward rapidity ($2.5 < y < 4.0$) the \j are measured via the $\mu^{+}\mu^{-}$ decay channel and at mid-rapidity ($|y| < 0.9$) via the $e^{+}e^{-}$ decay channel. The results are presented as a function of \pt\ in the range $0 < p_{\rm T} < 12$~GeV/$c$. For the dimuon channel different collision centralities are also investigated.

The ALICE detector is described in~\cite{Aamodt:2008zz}. 
At forward rapidity the production of quarkonia is measured with the muon spectrometer~\footnote{In the ALICE reference frame, the muon spectrometer covers a negative $\eta$ range and consequently a negative $y$ range.
We have chosen to present our results with a positive $y$ notation, due to the symmetry of the collision system.} consisting of a front absorber stopping the hadrons followed by five tracking stations comprising two planes of cathode pad chambers each, with the third station inside a dipole magnet. 
The tracking apparatus is completed by a triggering system made of four planes of resistive plate chambers downstream of an iron wall.
At mid-rapidity quarkonium production is measured with the central barrel detectors~\cite{Abelev:2012kr}. 
Tracking within $|\eta| < 0.9$ is performed by the Inner Tracking System (ITS)~\cite{Aamodt:2010aa} and the Time Projection Chamber (TPC)~\cite{Alme:2010ke}.
The specific ionization energy loss (\dEdx) in the gas of the TPC is used for particle identification (PID). 
In addition the Silicon Pixel Detector (SPD) is used to locate the interaction point. 
The SPD corresponds to the two innermost layers of the ITS covering respectively $|\eta| < 2.0$ and $|\eta| < 1.4$. 
The V0 counters~\cite{Abbas:2013taa}, consisting of two arrays of 32 scintillator sectors each covering $2.8 \leq \eta \leq 5.1$ (V0-A) and $-3.7 \leq \eta \leq -1.7$ (V0-C), are used as trigger and centrality detectors~\cite{Abelev:2013qoq,Adam:2015ptt}.
As described later, the SPD, TPC, V0-A, and V0-C are also used as event plane detectors.
All of these detectors have full azimuthal coverage.

The data were collected in 2015.
The analysis at mid-rapidity uses minimum bias (MB) Pb--Pb collisions.
The MB trigger requires a signal in both V0-A and V0-C and is fully efficient for the centrality range 0--90\%.
At forward rapidity, the analysis uses opposite-sign dimuon (MU) triggered Pb--Pb collisions.
The MU trigger requires a MB trigger and at least a pair of opposite-sign track segments in the muon trigger system, each with a \pt\ above the threshold of the on-line trigger algorithm, set to provide 50\% efficiency for muon tracks with $\pt = 1$ GeV/$c$. 
The beam-induced  background was further reduced offline using the V0 and the zero degree calorimeter (ZDC) timing information. 
The contribution from electromagnetic processes was removed by requiring a minimum energy deposited in the neutron ZDCs~\cite{ALICE:2012aa}. 
The resulting data samples correspond to integrated luminosities of about 13 $\mu$b$^{-1}$ and 225 $\mu$b$^{-1}$ at mid- and forward rapidity, respectively. 
 
\j candidates are formed by combining pairs of opposite-sign tracks reconstructed in the geometrical acceptance of the muon spectrometer or central barrel.
The reconstructed tracks in the muon tracker are required to match a track segment in the muon trigger system above the aforementioned \pt\ threshold.
At mid-rapidity the tracks must pass a \pt\ cut of $1$ GeV/$c$ and an electron selection criterion based on the expected \dEdx~\cite{Alme:2010ke}.

The dimuon \vv is calculated using event plane (EP) based methods. 
The angle of the reaction plane of the collision is estimated, event by event, by the second harmonic EP angle $\Psi$~\cite{Poskanzer:1998yz}, which is obtained from the azimuthal distribution of reconstructed tracks in the TPC or track segments in the SPD for the mid- and forward rapidity analyses, respectively. 
Effects of non-uniform acceptance in the EP determination are corrected using the methods described in~\cite{ Selyuzhenkov:2007zi}. 
At mid-rapidity, the EP was calculated for each electron pair subtracting the contribution of the pair tracks to remove auto-correlations. 

\begin{figure}[t!]
\includegraphics[width=0.95\linewidth]{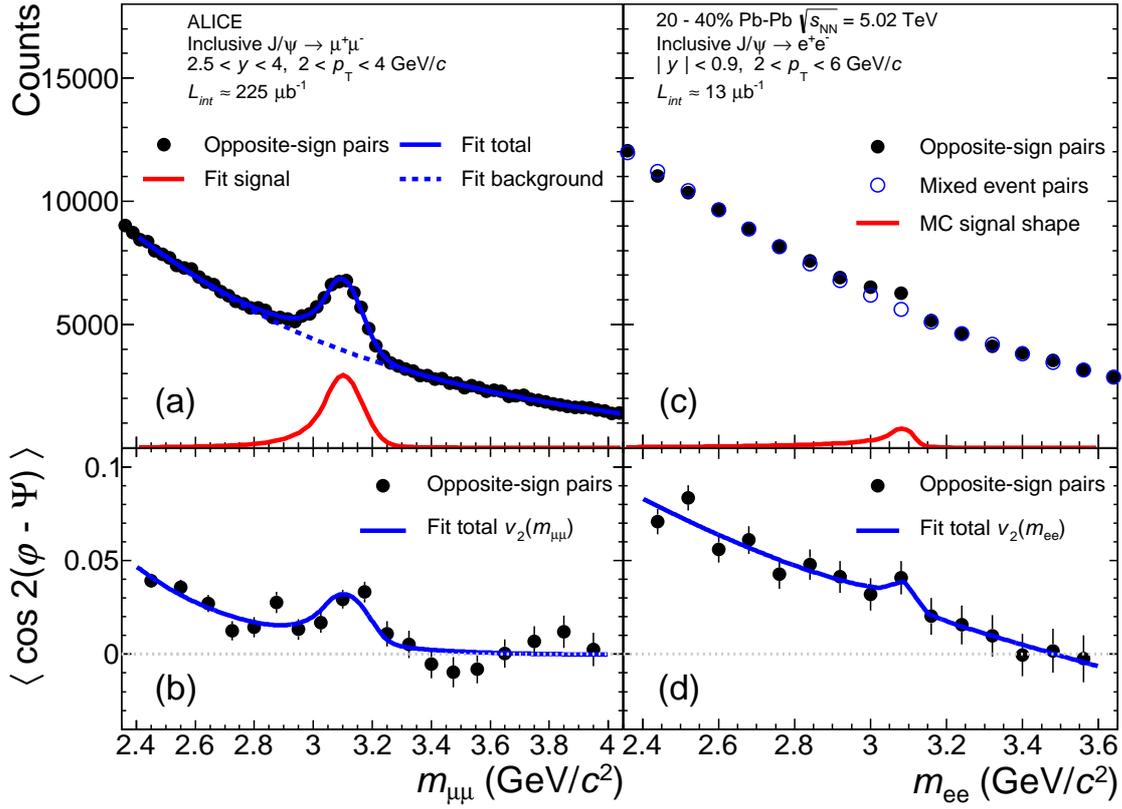}
\caption{\label{fig:massv2} (color online) Invariant mass distribution (top)  and $\langle \cos 2 (\varphi - \Psi) \rangle$ as a function of $m_{\ell\ell}$ (bottom) of opposite-sign dimuons (left) with $2 < p_{\rm T}<4$ GeV/$c$ and $2.5 < y < 4$ and dielectrons (right) with $2 < p_{\rm T} < 6$ GeV/$c$ and $|y| < 0.9$, in semi-central (20--40\%) Pb--Pb collisions.}
\end{figure}

The \j \pt\ results were obtained, as proposed in ~\cite{Borghini:2004ra}, by fitting the distribution of $v_{2} = \langle \cos 2 (\varphi - \Psi) \rangle$ versus the invariant mass ($m_{\ell\ell}$) of the dilepton pair, with $\varphi$ being its azimuthal angle.
The total flow $v_{2}(m_{\ell\ell})$ is the combination of the signal and the background flow and can be expressed as

\begin{equation}\label{eq:v2Tot}
v_{2}(m_{\ell\ell}) = v_{2}^{\rm sig}\alpha(m_{\ell\ell}) + v_{2}^{\rm bkg}(m_{\ell\ell})[1-\alpha(m_{\ell\ell})],
\end{equation}
where $v_{2}^{\rm sig}$ and $v_{2}^{\rm bkg}$ are the elliptic flow of the \j signal (S) and of the background (B), respectively (see bottom panels of Fig.~\ref{fig:massv2}).
The signal fraction $\alpha(m_{\ell\ell}) = S(m_{\ell\ell})/(S(m_{\ell\ell})+B(m_{\ell\ell}))$ was extracted from fits to the invariant mass distribution (see top panels of Fig.~\ref{fig:massv2}) in each \pt\ and centrality class.

At forward rapidity, the \j peak ($S$-term of $\alpha(m_{\ell\ell})$) is fit with an extended Crystal Ball function or a pseudo-Gaussian, both composed of a Gaussian core with non-Gaussian tails~\cite{ALICE-PUBLIC-2015-006}. The underlying continuum ($B$-term of $\alpha(m_{\ell\ell})$) is described with the ratio of second- to third-order polynomials, a pseudo-Gaussian with a width quadratically varying with mass, or Chebyshev polynomials of order six.
The background flow $v_{2}^{\rm bkg}$ was parametrized using a second-order polynomial, a Chebyshev polynomial of order four, or the product of a first order polynomial and an exponential function.
At mid rapidity, the underlying continuum was estimated combining opposite-sign electrons from different events (using an event-mixing technique) or combining same-sign electrons from the same event.
After removing the underlying continuum, the \j signal was obtained by counting the number of dielectrons or from a fit with a MC-generated shape.
The background flow was parametrized using a second-, third- or fifth-order polynomial depending on the \pt\ class.
Additionally, the PID and track-quality selection criteria were varied as part of the systematic uncertainty evaluation.

The \j \vv and its statistical uncertainty in each \pt\ and centrality class were determined as the average of the $v_{2}^{\rm sig}$ obtained by fitting $v_{2}(m_{\ell\ell})$ using Eq.~\ref{eq:v2Tot} with the various $\alpha(m_{\ell\ell})$ and $v_{2}^{\rm bkg}(m_{\ell\ell})$ parametrizations in several invariant mass ranges, while the corresponding systematic uncertainties were defined as the RMS of these results.
A similar method was used to extract the uncorrected (for detector acceptance and efficiency) average transverse momentum of the reconstructed \j in each centrality and \pt\ class, which is used to locate the data points when plotted as a function of \pt.
Consistent \vv values were obtained using an alternative method~\cite{Poskanzer:1998yz} in which the \j raw yield is extracted, as described before, in bins of $(\varphi-\Psi)$ and \pt\ is evaluated by fitting the data with the function $\frac{dN}{d(\varphi-\Psi)} = A[1 + 2v_{2}\cos 2(\varphi-\Psi)]$, where $A$ is a normalization constant.

Non-flow effects (J/$\psi$-EP correlations not related to the the initial geometry symmetry plane, such as higher-mass particle decays or jets) were estimated to be small with respect to the other uncertainties by repeating the analysis at forward rapidity using the EP determined in either V0-A ($\Delta \eta = 5.3$) or V0-C (no $\eta$ gap) detector.

The finite resolution in the EP determination smears out the azimuthal distributions and lowers the value of the measured anisotropy~\cite{Poskanzer:1998yz}.
The SPD- and TPC-based EP resolutions were determined by applying the 3 sub-event method~\cite{Poskanzer:1998yz}.
For the SPD (TPC), the 3 sub-events were obtained using V0-A, V0-C and SPD, with $\Delta \eta_{\rm V0A-SPD}=1.4$ ($\Delta \eta_{\rm V0A-TPC}=1.9$), $\Delta \eta_{\rm V0A-V0C}=4.5$ and $\Delta \eta_{\rm SPD-V0C}=0.3$ ($\Delta \eta_{\rm TPC-V0C}=0.8$) pseudo-rapidity gaps.
A systematic uncertainty  of $1\%$ on the EP determination was estimated exploiting the availability of different sub-events, built from the multiplicity measurement in the V0-A or V0-C, track segments in the SPD, and tracks in the TPC.
The EP resolution for each wide centrality class was calculated as the average of the values obtained in finer classes weighted by the number of reconstructed $J/\psi$.
Table~\ref{tab:res} shows the corresponding resolution for each centrality class, applied to the forward rapidity results. 
For the mid-rapidity result, the TPC EP resolution is $0.880 \pm 0.009$ (syst) in the centrality class 20--40\%.

\begin{table}[h]
\begin{center}
\begin{tabular}{ccc}
\hline
\hline
Centrality  & $\langle N_{\rm part} \rangle$  &  EP resolution   \\
\hline
  5--20\% & 287 $\pm$ 4 & $0.873 \pm 0.009$ \\
20--40\% & 160 $\pm$ 3 & $0.910 \pm 0.009$ \\
40--60\% &   70 $\pm$ 2 & $0.832 \pm 0.008$ \\
\hline
\hline
\end{tabular}
\end{center}
\caption{\label{tab:res} Average number of participants $\langle N_{\rm{part}} \rangle$ and SPD EP resolution for each centrality class (expressed in percentage of the nuclear cross section)~\cite{Adam:2015ptt}. The quoted uncertainties are systematic.}
\end{table}

At forward rapidity, the \j reconstruction efficiency depends on the detector occupancy, which could bias the \vv measurement.
This effect was evaluated by embedding azimuthally isotropic simulated decays into real events.
The resulting \vv does not deviate from zero by more than 0.006 in the centrality and \pt\ classes considered.
This value is used as a conservative systematic uncertainty on all measured \vv values.

\begin{figure}[t!]
\includegraphics[width=0.95\linewidth]{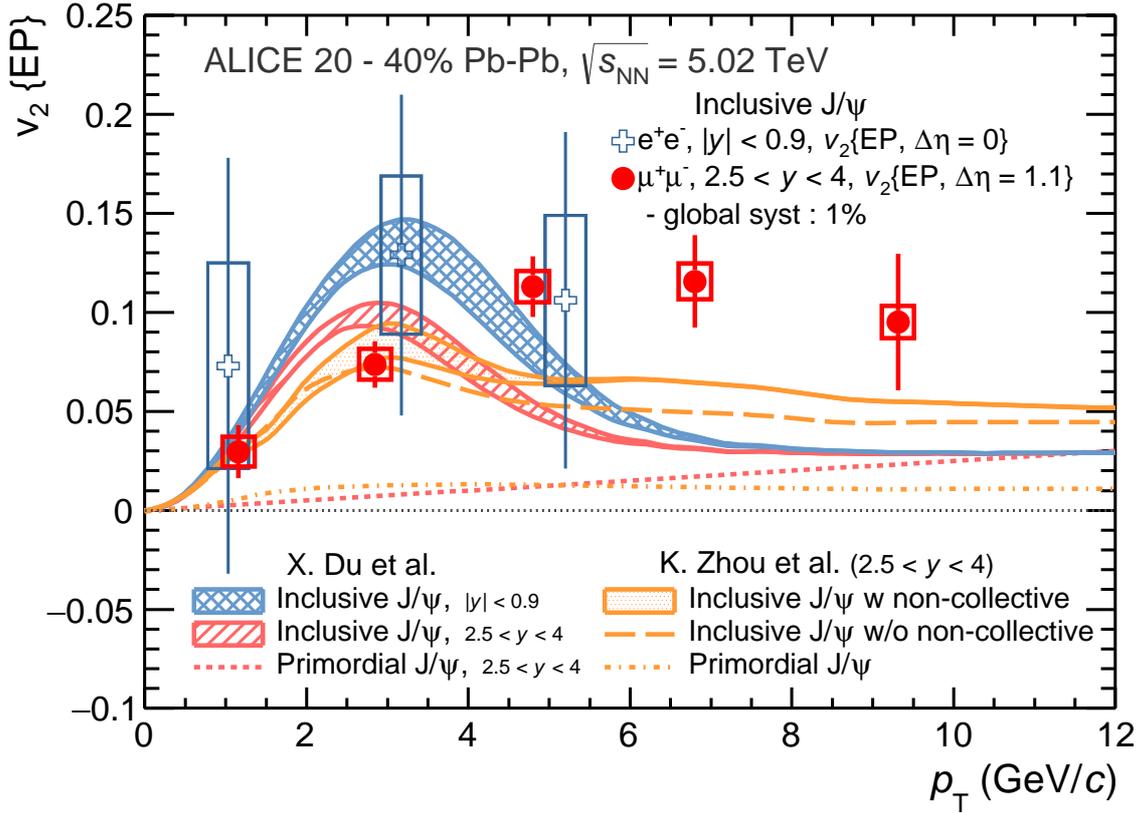}
\caption{\label{fig:v2VsTheory} (color online) Inclusive \j \vpt\ at forward and mid-rapidity for semi-central (20--40\%) \pb collisions at \snnfive~ TeV. 
Calculations from transports model by~\cite{Du:2015wha} and~\cite{Zhou:2014kka} are also shown.}
\end{figure}

Figure \ref{fig:v2VsTheory} shows \j \vpt\ at forward and mid-rapidity in semi-central (20--40\%) \pb collisions at \snnfive~ TeV. 
The \pt\ ranges are 0--2, 2--4, 4--6, 6--8, and 8--12 GeV/$c$ and 0--2, 2--6, and 4--12 GeV/$c$ at forward and mid-rapidity, respectively.  
The vertical bars indicate the statistical uncertainties, while the boxes indicate the uncorrelated systematic uncertainties.
The global relative systematic uncertainty on the EP resolution is 1.0\% and is correlated with \pt. 
At forward rapidity, a positive \vv is observed for semi-central collisions (20--40\%). 
Including statistical and systematic uncertainties the significance of a non-zero \vv is as large as $6.6 \sigma$ in the \pt\ class 4--6 GeV/$c$. 
The \j \vv increases with \pt\ up to $v_{2}=0.113\pm0.015(\rm{stat})\pm0.008(\rm{syst})$ at $4 < p_{\mathrm{T}} < 6$ GeV/$c$.
The \j \vpt\ at mid-rapidity is similar to that at forward rapidity, albeit with large uncertainties.
At mid-rapidity, the \j \vv in the range $2 < p_{\mathrm{T}} <  6$ GeV/$c$ is $v_{2} = 0.129\pm0.080(\rm{stat})\pm0.040(\rm{syst})$.

Transport model calculations including a large \j (re)generation component (about 50\% for semi-central collisions) from deconfined charm quarks in the medium~\cite{Du:2015wha,Zhao:2012gc,Zhou:2014kka} are also shown in Fig.~\ref{fig:v2VsTheory}. 
In the model by Du \emph{et al.}~\cite{Du:2015wha} (TM1) the \vv of inclusive \j (hashed and double-hashed bands at forward and mid-rapidity) has three origins.
First, thermalized charm quarks in the medium transfer a significant elliptic flow to (re)generated J/$\psi$.
Second, primordial \j traverse a longer path through the medium when emitted out-of-plane than in-plane resulting in a small apparent $v_{2}$  (pair dissociation by interactions with the surrounding color charges).
Third, when the b quarks thermalize their flow will be transferred to b-hadrons at hadronization and to non-prompt \j from the b-hadron decay.
The second component (survival provability of primordial J/$\psi$) is represented as a short-dashed line to highlight the small \j \vv in the absence of heavy-quark collective flow.  
The model by Zhou \emph{et al.}~\cite{Zhou:2014kka} (TM2) includes an additional non-collective \j \vv component, which arises from the modification of the quarkonium production in the presence of a strong magnetic field in the early stage of the heavy-ion collision~\cite{Guo:2015nsa}. 
The calculations of TM2 are shown at forward rapidity with (shaded band) and without (long-dashed line) the non-collective \j \vv component. 
As for TM1, the \vv resulting from the different in-plane than out-of-plane survival probability of primordial \j is shown as a dash-dotted line.  

TM1~\cite{Du:2015wha} is able to describe qualitatively the \j \raa\ measurements by ALICE reported in~\cite{Adam:2016rdg}. 
The model also agrees with ALICE \j \vv measurements at forward rapidity at \snntwo~ TeV~\cite{ALICE:2013xna} and at mid-rapidity at \snnfive~ TeV. 
However, at high \pt\ ($p_{\rm T} > 4$ GeV/$c$), clear discrepancies are observed between the model and the \j \vv at forward rapidity and \snnfive~ TeV. 
Some tension is also seen between the calculations of this model and the \raa\ measurement by ALICE in this higher \pt\ range in~\cite{Adam:2016rdg}. 
At lower \pt\, the model reproduces the magnitude of the measurement by a dominant contribution of \j elliptic flow inherited from thermalized charm quarks. 
However, the overall shape of the \vpt\ is missed and the \vv at high \pt\ is underestimated. 
This disagreement suggests a missing mechanism in the model. 
Similar conclusions can be derived from the comparison to TM2~\cite{Zhou:2014kka}. 
The addition of the \vv arising from a possible strong magnetic field in the early stage of heavy-ion collisions~\cite{Guo:2015nsa} improves the comparison with the measured \j \vv at forward rapidity, especially at high \pt. 
Such non-collective component was able to reproduce the prompt \j \vv at high \pt\ measured by CMS in \pb collisions at \snntwo ~ TeV~\cite{Khachatryan:2016ypw}.

\begin{figure}[t!]
\includegraphics[width=0.95\linewidth]{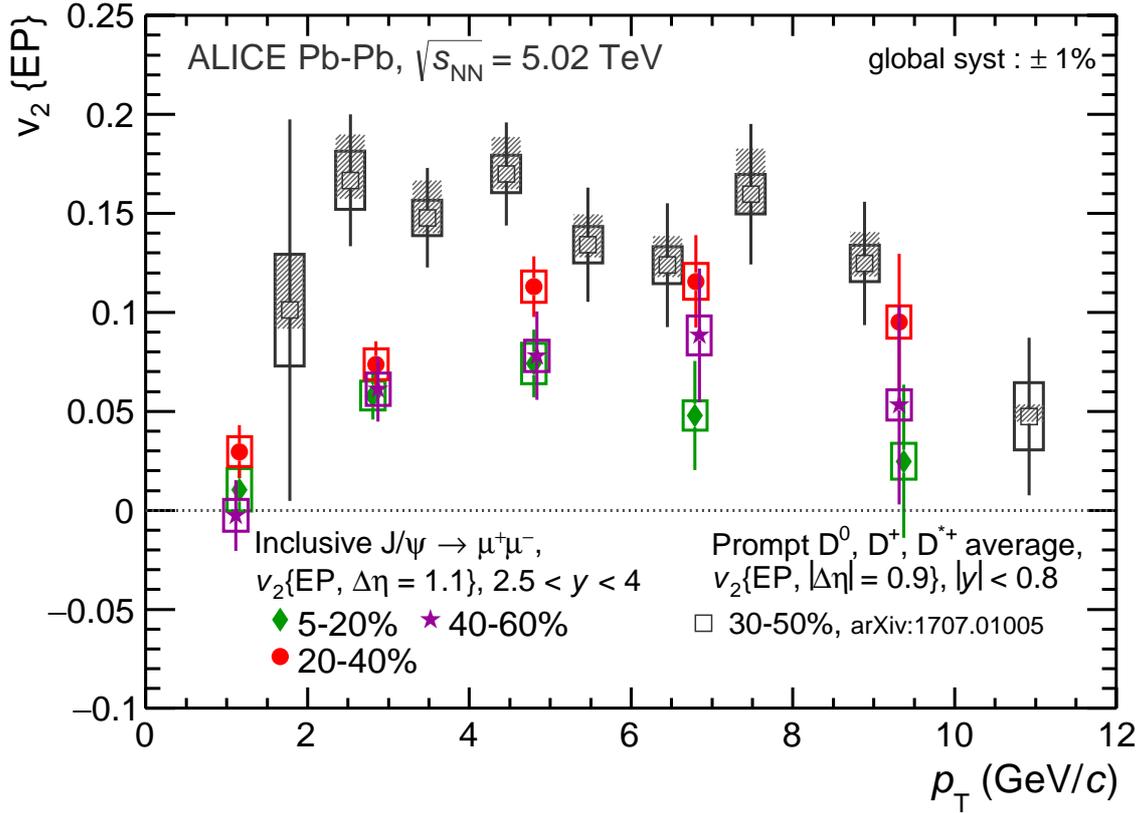}
\caption{\label{fig:v2pt} (color online) Inclusive \j \vpt\ at forward rapidity in \pb collisions at \snnfive~ TeV for three centrality classes,  5--20\%, 20--40\%, and 40--60\%.  The average of ${\rm D^0}$, ${\rm D^+}$ and ${\rm D^{*+}}$ \vpt\ at mid-y in the centrality class 30--50\% is also shown for comparison~\cite{Acharya:2017qps}.}
\end{figure}

Figure~\ref{fig:v2pt} presents the \pt\ dependence of the \j \vv at forward rapidity for three centrality classes, 5--20\%, 20--40\%, and 40--60\%. 
As in semi-central (20--40\%) collisions, a significant \vv is also observed for \j with  $2 <  p_{\mathrm{T}} < 8$ GeV/$c$ in the 5--20\% and 40--60\% centrality classes.
The \pt\ dependence of the \j \vv at forward rapidity is consistent within uncertainties in the three centrality classes presented here.
The \j \vpt\ appears to be maximum for the 20--40\% centrality class and tends to decrease for more central or peripheral collisions.  
Interestingly, for identified light hadrons in \pb collisions at \snntwo~ TeV, the \vpt\ is maximum in the 40--60\% centrality class and decreases for more central collisions~\cite{Abelev:2014pua}.
This different behavior could be understood in the framework of transport models by the increasing contribution of \j regeneration for more central collisions~\cite{Du:2015wha,Zhao:2012gc}.

Also shown in Fig.~\ref{fig:v2pt} is the \vpt\ of prompt D-mesons in \pb collisions at  \snnfive~ TeV for the 30--50\% centrality class measured by ALICE at mid-rapidity~\cite{Acharya:2017qps}. 
The vertical bars indicate the statistical uncertainties, the open boxes the  uncorrelated systematic uncertainties and the shaded boxes the feed-down uncertainties.
Although the centrality and rapidity ranges are different, it is clear that at low \pt\ ($\pt < 4$ GeV/$c$) the \vv of D mesons is higher than that of \j mesons.
The large values of the measured \vv of both D and \j mesons support the conclusion that both D and \j mesons inherit their flow from thermalized charm quarks.

In summary, we report the ALICE measurements of inclusive \j elliptic flow at forward and mid-rapidity in \pb collisions at \snnfive~ TeV. 
At forward rapidity, the \pt\ dependence of the \j \vv was measured in the 5--20\%, 20--40\%, and 40--60\% centrality classes for $p_{\rm T} < 12$ GeV/$c$. 
For all the reported centrality classes a significant \j \vv signal is observed in the intermediate region $2 < p_{\rm T} < 8$ GeV/$c$. 
The results unambiguously establish for the first time that \j mesons exhibit collective flow. At mid-rapidity, the \pt\ dependence of the \j \vv was measured in semi-central 20--40\% collisions and is found to be similar to the measurement at forward rapidity, albeit with larger uncertainties.
At high \pt, transport models underestimate the measured \j $v_{2}$. 
The origin of such discrepancy is currently not understood and suggests a missing mechanism in the models.
At low \pt, the magnitude of the observed \vv is achieved within transport models implementing a strong \j (re)generation component from (re)combination of thermalized charm quarks in the QGP. 
Thus, the measurement of the \j elliptic flow combined with the \raa\ provides substantial evidence for thermalized charm quarks and (re)generation of J/$\psi$.
               

%
\newenvironment{acknowledgement}{\relax}{\relax}
\begin{acknowledgement}
\section*{Acknowledgements}

The ALICE Collaboration would like to thank all its engineers and technicians for their invaluable contributions to the construction of the experiment and the CERN accelerator teams for the outstanding performance of the LHC complex.
The ALICE Collaboration gratefully acknowledges the resources and support provided by all Grid centres and the Worldwide LHC Computing Grid (WLCG) collaboration.
The ALICE Collaboration acknowledges the following funding agencies for their support in building and running the ALICE detector:
A. I. Alikhanyan National Science Laboratory (Yerevan Physics Institute) Foundation (ANSL), State Committee of Science and World Federation of Scientists (WFS), Armenia;
Austrian Academy of Sciences and Nationalstiftung f\"{u}r Forschung, Technologie und Entwicklung, Austria;
Ministry of Communications and High Technologies, National Nuclear Research Center, Azerbaijan;
Conselho Nacional de Desenvolvimento Cient\'{\i}fico e Tecnol\'{o}gico (CNPq), Universidade Federal do Rio Grande do Sul (UFRGS), Financiadora de Estudos e Projetos (Finep) and Funda\c{c}\~{a}o de Amparo \`{a} Pesquisa do Estado de S\~{a}o Paulo (FAPESP), Brazil;
Ministry of Science \& Technology of China (MSTC), National Natural Science Foundation of China (NSFC) and Ministry of Education of China (MOEC) , China;
Ministry of Science, Education and Sport and Croatian Science Foundation, Croatia;
Ministry of Education, Youth and Sports of the Czech Republic, Czech Republic;
The Danish Council for Independent Research | Natural Sciences, the Carlsberg Foundation and Danish National Research Foundation (DNRF), Denmark;
Helsinki Institute of Physics (HIP), Finland;
Commissariat \`{a} l'Energie Atomique (CEA) and Institut National de Physique Nucl\'{e}aire et de Physique des Particules (IN2P3) and Centre National de la Recherche Scientifique (CNRS), France;
Bundesministerium f\"{u}r Bildung, Wissenschaft, Forschung und Technologie (BMBF) and GSI Helmholtzzentrum f\"{u}r Schwerionenforschung GmbH, Germany;
General Secretariat for Research and Technology, Ministry of Education, Research and Religions, Greece;
National Research, Development and Innovation Office, Hungary;
Department of Atomic Energy Government of India (DAE), Department of Science and Technology, Government of India (DST), University Grants Commission, Government of India (UGC) and Council of Scientific and Industrial Research (CSIR), India;
Indonesian Institute of Science, Indonesia;
Centro Fermi - Museo Storico della Fisica e Centro Studi e Ricerche Enrico Fermi and Istituto Nazionale di Fisica Nucleare (INFN), Italy;
Institute for Innovative Science and Technology , Nagasaki Institute of Applied Science (IIST), Japan Society for the Promotion of Science (JSPS) KAKENHI and Japanese Ministry of Education, Culture, Sports, Science and Technology (MEXT), Japan;
Consejo Nacional de Ciencia (CONACYT) y Tecnolog\'{i}a, through Fondo de Cooperaci\'{o}n Internacional en Ciencia y Tecnolog\'{i}a (FONCICYT) and Direcci\'{o}n General de Asuntos del Personal Academico (DGAPA), Mexico;
Nederlandse Organisatie voor Wetenschappelijk Onderzoek (NWO), Netherlands;
The Research Council of Norway, Norway;
Commission on Science and Technology for Sustainable Development in the South (COMSATS), Pakistan;
Pontificia Universidad Cat\'{o}lica del Per\'{u}, Peru;
Ministry of Science and Higher Education and National Science Centre, Poland;
Korea Institute of Science and Technology Information and National Research Foundation of Korea (NRF), Republic of Korea;
Ministry of Education and Scientific Research, Institute of Atomic Physics and Romanian National Agency for Science, Technology and Innovation, Romania;
Joint Institute for Nuclear Research (JINR), Ministry of Education and Science of the Russian Federation and National Research Centre Kurchatov Institute, Russia;
Ministry of Education, Science, Research and Sport of the Slovak Republic, Slovakia;
National Research Foundation of South Africa, South Africa;
Centro de Aplicaciones Tecnol\'{o}gicas y Desarrollo Nuclear (CEADEN), Cubaenerg\'{\i}a, Cuba, Ministerio de Ciencia e Innovacion and Centro de Investigaciones Energ\'{e}ticas, Medioambientales y Tecnol\'{o}gicas (CIEMAT), Spain;
Swedish Research Council (VR) and Knut \& Alice Wallenberg Foundation (KAW), Sweden;
European Organization for Nuclear Research, Switzerland;
National Science and Technology Development Agency (NSDTA), Suranaree University of Technology (SUT) and Office of the Higher Education Commission under NRU project of Thailand, Thailand;
Turkish Atomic Energy Agency (TAEK), Turkey;
National Academy of  Sciences of Ukraine, Ukraine;
Science and Technology Facilities Council (STFC), United Kingdom;
National Science Foundation of the United States of America (NSF) and United States Department of Energy, Office of Nuclear Physics (DOE NP), United States of America.
\end{acknowledgement}

\bibliographystyle{utphys}
\bibliography{Jpsiv2PbPb15_CernPreprint}

\newpage
\appendix
\section{The ALICE Collaboration}
\label{app:collab}

\begingroup
\small
\begin{flushleft}
S.~Acharya\Irefn{org137}\And 
D.~Adamov\'{a}\Irefn{org94}\And 
J.~Adolfsson\Irefn{org34}\And 
M.M.~Aggarwal\Irefn{org99}\And 
G.~Aglieri Rinella\Irefn{org35}\And 
M.~Agnello\Irefn{org31}\And 
N.~Agrawal\Irefn{org48}\And 
Z.~Ahammed\Irefn{org137}\And 
S.U.~Ahn\Irefn{org79}\And 
S.~Aiola\Irefn{org141}
% ...
\And
\renewcommand\labelenumi{\textsuperscript{\theenumi}~}

\section*{Affiliation notes}
\renewcommand\theenumi{\roman{enumi}}
\begin{Authlist}
\item \Adef{org*}Deceased
\item \Adef{orgI}Dipartimento DET del Politecnico di Torino, Turin, Italy
\item \Adef{orgII}M.V. Lomonosov Moscow State University, D.V. Skobeltsyn Institute of Nuclear, Physics, Moscow, Russia
\item \Adef{orgIII}Department of Applied Physics, Aligarh Muslim University, Aligarh, India
\item \Adef{orgIV}Institute of Theoretical Physics, University of Wroclaw, Poland
\end{Authlist}

\section*{Collaboration Institutes}
\renewcommand\theenumi{\arabic{enumi}~}
\begin{Authlist}
\item \Idef{org1}A.I. Alikhanyan National Science Laboratory (Yerevan Physics Institute) Foundation, Yerevan, Armenia
\item \Idef{org2}Benem\'{e}rita Universidad Aut\'{o}noma de Puebla, Puebla, Mexico
\item \Idef{org3}Bogolyubov Institute for Theoretical Physics, Kiev, Ukraine
\item \Idef{org4}Bose Institute, Department of Physics  and Centre for Astroparticle Physics and Space Science (CAPSS), Kolkata, India
\item \Idef{org5}Budker Institute for Nuclear Physics, Novosibirsk, Russia
\item \Idef{org6}California Polytechnic State University, San Luis Obispo, California, United States
\item \Idef{org7}Central China Normal University, Wuhan, China
\item \Idef{org8}Centre de Calcul de l'IN2P3, Villeurbanne, Lyon, France
\item \Idef{org9}Centro de Aplicaciones Tecnol\'{o}gicas y Desarrollo Nuclear (CEADEN), Havana, Cuba
\item \Idef{org10}Centro de Investigaciones Energ\'{e}ticas Medioambientales y Tecnol\'{o}gicas (CIEMAT), Madrid, Spain
\item \Idef{org11}Centro de Investigaci\'{o}n y de Estudios Avanzados (CINVESTAV), Mexico City and M\'{e}rida, Mexico
\item \Idef{org12}Centro Fermi - Museo Storico della Fisica e Centro Studi e Ricerche ``Enrico Fermi', Rome, Italy
\item \Idef{org13}Chicago State University, Chicago, Illinois, United States
\item \Idef{org14}China Institute of Atomic Energy, Beijing, China
\item \Idef{org15}COMSATS Institute of Information Technology (CIIT), Islamabad, Pakistan
\item \Idef{org16}Departamento de F\'{\i}sica de Part\'{\i}culas and IGFAE, Universidad de Santiago de Compostela, Santiago de Compostela, Spain
\item \Idef{org17}Department of Physics, Aligarh Muslim University, Aligarh, India
\item \Idef{org18}Department of Physics, Ohio State University, Columbus, Ohio, United States
\item \Idef{org19}Department of Physics, Pusan National University, Pusan, Republic of Korea
\item \Idef{org20}Department of Physics, Sejong University, Seoul, Republic of Korea
\item \Idef{org21}Department of Physics, University of Oslo, Oslo, Norway
\item \Idef{org22}Department of Physics and Technology, University of Bergen, Bergen, Norway
\item \Idef{org23}Dipartimento di Fisica dell'Universit\`{a} 'La Sapienza' and Sezione INFN, Rome, Italy
\item \Idef{org24}Dipartimento di Fisica dell'Universit\`{a} and Sezione INFN, Cagliari, Italy
\item \Idef{org25}Dipartimento di Fisica dell'Universit\`{a} and Sezione INFN, Trieste, Italy
\item \Idef{org26}Dipartimento di Fisica dell'Universit\`{a} and Sezione INFN, Turin, Italy
\item \Idef{org27}Dipartimento di Fisica e Astronomia dell'Universit\`{a} and Sezione INFN, Bologna, Italy
\item \Idef{org28}Dipartimento di Fisica e Astronomia dell'Universit\`{a} and Sezione INFN, Catania, Italy
\item \Idef{org29}Dipartimento di Fisica e Astronomia dell'Universit\`{a} and Sezione INFN, Padova, Italy
\item \Idef{org30}Dipartimento di Fisica `E.R.~Caianiello' dell'Universit\`{a} and Gruppo Collegato INFN, Salerno, Italy
\item \Idef{org31}Dipartimento DISAT del Politecnico and Sezione INFN, Turin, Italy
\item \Idef{org32}Dipartimento di Scienze e Innovazione Tecnologica dell'Universit\`{a} del Piemonte Orientale and INFN Sezione di Torino, Alessandria, Italy
\item \Idef{org33}Dipartimento Interateneo di Fisica `M.~Merlin' and Sezione INFN, Bari, Italy
\item \Idef{org34}Division of Experimental High Energy Physics, University of Lund, Lund, Sweden
\item \Idef{org35}European Organization for Nuclear Research (CERN), Geneva, Switzerland
\item \Idef{org36}Excellence Cluster Universe, Technische Universit\"{a}t M\"{u}nchen, Munich, Germany
\item \Idef{org37}Faculty of Engineering, Bergen University College, Bergen, Norway
\item \Idef{org38}Faculty of Mathematics, Physics and Informatics, Comenius University, Bratislava, Slovakia
\item \Idef{org39}Faculty of Nuclear Sciences and Physical Engineering, Czech Technical University in Prague, Prague, Czech Republic
\item \Idef{org40}Faculty of Science, P.J.~\v{S}af\'{a}rik University, Ko\v{s}ice, Slovakia
\item \Idef{org41}Faculty of Technology, Buskerud and Vestfold University College, Tonsberg, Norway
\item \Idef{org42}Frankfurt Institute for Advanced Studies, Johann Wolfgang Goethe-Universit\"{a}t Frankfurt, Frankfurt, Germany
\item \Idef{org43}Gangneung-Wonju National University, Gangneung, Republic of Korea
\item \Idef{org44}Gauhati University, Department of Physics, Guwahati, India
\item \Idef{org45}Helmholtz-Institut f\"{u}r Strahlen- und Kernphysik, Rheinische Friedrich-Wilhelms-Universit\"{a}t Bonn, Bonn, Germany
\item \Idef{org46}Helsinki Institute of Physics (HIP), Helsinki, Finland
\item \Idef{org47}Hiroshima University, Hiroshima, Japan
\item \Idef{org48}Indian Institute of Technology Bombay (IIT), Mumbai, India
\item \Idef{org49}Indian Institute of Technology Indore, Indore, India
\item \Idef{org50}Indonesian Institute of Sciences, Jakarta, Indonesia
\item \Idef{org51}INFN, Laboratori Nazionali di Frascati, Frascati, Italy
\item \Idef{org52}INFN, Sezione di Bari, Bari, Italy
\item \Idef{org53}INFN, Sezione di Bologna, Bologna, Italy
\item \Idef{org54}INFN, Sezione di Cagliari, Cagliari, Italy
\item \Idef{org55}INFN, Sezione di Catania, Catania, Italy
\item \Idef{org56}INFN, Sezione di Padova, Padova, Italy
\item \Idef{org57}INFN, Sezione di Roma, Rome, Italy
\item \Idef{org58}INFN, Sezione di Torino, Turin, Italy
\item \Idef{org59}INFN, Sezione di Trieste, Trieste, Italy
\item \Idef{org60}Inha University, Incheon, Republic of Korea
\item \Idef{org61}Institut de Physique Nucl\'eaire d'Orsay (IPNO), Universit\'e Paris-Sud, CNRS-IN2P3, Orsay, France
\item \Idef{org62}Institute for Nuclear Research, Academy of Sciences, Moscow, Russia
\item \Idef{org63}Institute for Subatomic Physics of Utrecht University, Utrecht, Netherlands
\item \Idef{org64}Institute for Theoretical and Experimental Physics, Moscow, Russia
\item \Idef{org65}Institute of Experimental Physics, Slovak Academy of Sciences, Ko\v{s}ice, Slovakia
\item \Idef{org66}Institute of Physics, Academy of Sciences of the Czech Republic, Prague, Czech Republic
\item \Idef{org67}Institute of Physics, Bhubaneswar, India
\item \Idef{org68}Institute of Space Science (ISS), Bucharest, Romania
\item \Idef{org69}Institut f\"{u}r Informatik, Johann Wolfgang Goethe-Universit\"{a}t Frankfurt, Frankfurt, Germany
\item \Idef{org70}Institut f\"{u}r Kernphysik, Johann Wolfgang Goethe-Universit\"{a}t Frankfurt, Frankfurt, Germany
\item \Idef{org71}Institut f\"{u}r Kernphysik, Westf\"{a}lische Wilhelms-Universit\"{a}t M\"{u}nster, M\"{u}nster, Germany
\item \Idef{org72}Instituto de Ciencias Nucleares, Universidad Nacional Aut\'{o}noma de M\'{e}xico, Mexico City, Mexico
\item \Idef{org73}Instituto de F\'{i}sica, Universidade Federal do Rio Grande do Sul (UFRGS), Porto Alegre, Brazil
\item \Idef{org74}Instituto de F\'{\i}sica, Universidad Nacional Aut\'{o}noma de M\'{e}xico, Mexico City, Mexico
\item \Idef{org75}IRFU, CEA, Universit\'{e} Paris-Saclay, Saclay, France
\item \Idef{org76}iThemba LABS, National Research Foundation, Somerset West, South Africa
\item \Idef{org77}Joint Institute for Nuclear Research (JINR), Dubna, Russia
\item \Idef{org78}Konkuk University, Seoul, Republic of Korea
\item \Idef{org79}Korea Institute of Science and Technology Information, Daejeon, Republic of Korea
\item \Idef{org80}KTO Karatay University, Konya, Turkey
\item \Idef{org81}Laboratoire de Physique Subatomique et de Cosmologie, Universit\'{e} Grenoble-Alpes, CNRS-IN2P3, Grenoble, France
\item \Idef{org82}Lawrence Berkeley National Laboratory, Berkeley, California, United States
\item \Idef{org83}Moscow Engineering Physics Institute, Moscow, Russia
\item \Idef{org84}Nagasaki Institute of Applied Science, Nagasaki, Japan
\item \Idef{org85}National and Kapodistrian University of Athens, Physics Department, Athens, Greece
\item \Idef{org86}National Centre for Nuclear Studies, Warsaw, Poland
\item \Idef{org87}National Institute for Physics and Nuclear Engineering, Bucharest, Romania
\item \Idef{org88}National Institute of Science Education and Research, HBNI, Jatni, India
\item \Idef{org89}National Nuclear Research Center, Baku, Azerbaijan
\item \Idef{org90}National Research Centre Kurchatov Institute, Moscow, Russia
\item \Idef{org91}Niels Bohr Institute, University of Copenhagen, Copenhagen, Denmark
\item \Idef{org92}Nikhef, Nationaal instituut voor subatomaire fysica, Amsterdam, Netherlands
\item \Idef{org93}Nuclear Physics Group, STFC Daresbury Laboratory, Daresbury, United Kingdom
\item \Idef{org94}Nuclear Physics Institute, Academy of Sciences of the Czech Republic, \v{R}e\v{z} u Prahy, Czech Republic
\item \Idef{org95}Oak Ridge National Laboratory, Oak Ridge, Tennessee, United States
\item \Idef{org96}Petersburg Nuclear Physics Institute, Gatchina, Russia
\item \Idef{org97}Physics Department, Creighton University, Omaha, Nebraska, United States
\item \Idef{org98}Physics department, Faculty of science, University of Zagreb, Zagreb, Croatia
\item \Idef{org99}Physics Department, Panjab University, Chandigarh, India
\item \Idef{org100}Physics Department, University of Cape Town, Cape Town, South Africa
\item \Idef{org101}Physics Department, University of Jammu, Jammu, India
\item \Idef{org102}Physics Department, University of Rajasthan, Jaipur, India
\item \Idef{org103}Physikalisches Institut, Eberhard Karls Universit\"{a}t T\"{u}bingen, T\"{u}bingen, Germany
\item \Idef{org104}Physikalisches Institut, Ruprecht-Karls-Universit\"{a}t Heidelberg, Heidelberg, Germany
\item \Idef{org105}Physik Department, Technische Universit\"{a}t M\"{u}nchen, Munich, Germany
\item \Idef{org106}Research Division and ExtreMe Matter Institute EMMI, GSI Helmholtzzentrum f\"ur Schwerionenforschung GmbH, Darmstadt, Germany
\item \Idef{org107}Rudjer Bo\v{s}kovi\'{c} Institute, Zagreb, Croatia
\item \Idef{org108}Russian Federal Nuclear Center (VNIIEF), Sarov, Russia
\item \Idef{org109}Saha Institute of Nuclear Physics, Kolkata, India
\item \Idef{org110}School of Physics and Astronomy, University of Birmingham, Birmingham, United Kingdom
\item \Idef{org111}Secci\'{o}n F\'{\i}sica, Departamento de Ciencias, Pontificia Universidad Cat\'{o}lica del Per\'{u}, Lima, Peru
\item \Idef{org112}SSC IHEP of NRC Kurchatov institute, Protvino, Russia
\item \Idef{org113}Stefan Meyer Institut f\"{u}r Subatomare Physik (SMI), Vienna, Austria
\item \Idef{org114}SUBATECH, IMT Atlantique, Universit\'{e} de Nantes, CNRS-IN2P3, Nantes, France
\item \Idef{org115}Suranaree University of Technology, Nakhon Ratchasima, Thailand
\item \Idef{org116}Technical University of Ko\v{s}ice, Ko\v{s}ice, Slovakia
\item \Idef{org117}Technical University of Split FESB, Split, Croatia
\item \Idef{org118}The Henryk Niewodniczanski Institute of Nuclear Physics, Polish Academy of Sciences, Cracow, Poland
\item \Idef{org119}The University of Texas at Austin, Physics Department, Austin, Texas, United States
\item \Idef{org120}Universidad Aut\'{o}noma de Sinaloa, Culiac\'{a}n, Mexico
\item \Idef{org121}Universidade de S\~{a}o Paulo (USP), S\~{a}o Paulo, Brazil
\item \Idef{org122}Universidade Estadual de Campinas (UNICAMP), Campinas, Brazil
\item \Idef{org123}Universidade Federal do ABC, Santo Andre, Brazil
\item \Idef{org124}University of Houston, Houston, Texas, United States
\item \Idef{org125}University of Jyv\"{a}skyl\"{a}, Jyv\"{a}skyl\"{a}, Finland
\item \Idef{org126}University of Liverpool, Liverpool, United Kingdom
\item \Idef{org127}University of Tennessee, Knoxville, Tennessee, United States
\item \Idef{org128}University of the Witwatersrand, Johannesburg, South Africa
\item \Idef{org129}University of Tokyo, Tokyo, Japan
\item \Idef{org130}University of Tsukuba, Tsukuba, Japan
\item \Idef{org131}Universit\'{e} Clermont Auvergne, CNRS/IN2P3, LPC, Clermont-Ferrand, France
\item \Idef{org132}Universit\'{e} de Lyon, Universit\'{e} Lyon 1, CNRS/IN2P3, IPN-Lyon, Villeurbanne, Lyon, France
\item \Idef{org133}Universit\'{e} de Strasbourg, CNRS, IPHC UMR 7178, F-67000 Strasbourg, France, Strasbourg, France
\item \Idef{org134}Universit\`{a} degli Studi di Pavia, Pavia, Italy
\item \Idef{org135}Universit\`{a} di Brescia, Brescia, Italy
\item \Idef{org136}V.~Fock Institute for Physics, St. Petersburg State University, St. Petersburg, Russia
\item \Idef{org137}Variable Energy Cyclotron Centre, Kolkata, India
\item \Idef{org138}Warsaw University of Technology, Warsaw, Poland
\item \Idef{org139}Wayne State University, Detroit, Michigan, United States
\item \Idef{org140}Wigner Research Centre for Physics, Hungarian Academy of Sciences, Budapest, Hungary
\item \Idef{org141}Yale University, New Haven, Connecticut, United States
\item \Idef{org142}Yonsei University, Seoul, Republic of Korea
\item \Idef{org143}Zentrum f\"{u}r Technologietransfer und Telekommunikation (ZTT), Fachhochschule Worms, Worms, Germany
\end{Authlist}
\endgroup
\end{document}